\documentclass[twocolumn,aps,floatfix,superscriptaddress]{revtex4}
\pdfoutput=1 
\usepackage{amsmath,amssymb,eucal,graphicx}
\usepackage{color}

\begin{document}
\title{Trapping and Escape in a Turbid Medium}

\author{P. L. Krapivsky}
\affiliation{Department of Physics, Boston University, Boston, MA 02215, USA}
\author{S. Redner}
\affiliation{Santa Fe Institute, 1399 Hyde Park Road, Santa Fe, NM 87501, USA}

\begin{abstract}

  We investigate the absorption of diffusing molecules in a fluid-filled
  spherical beaker that contains many small reactive traps.  The molecules
  are absorbed either by hitting a trap or by escaping via the beaker walls.
  In the physical situation where the number $N$ of traps is large and their
  radii $a$ are small compared to the beaker radius $R$, the fraction of
  molecules $E$ that escape to the beaker wall and the complementary fraction
  $T$ that eventually are absorbed by the traps depend only on the
  dimensionless parameter combination $\lambda = Na/R$.  We compute $E$ and
  $T$ as a function of $\lambda$ for a spherical beaker and for beakers of
  other three-dimensional shapes. The asymptotic behavior is found to be
  universal: $1- E\sim \lambda$ for $\lambda\to 0$ and $E\sim\lambda^{-1/2}$
  for $\lambda\to\infty$.
\end{abstract}

\maketitle

\section{Introduction}

Consider a beaker filled with turbid medium---a fluid that contains many
small reactive traps.  We assume that the trap concentration is sufficiently
low that interactions between traps can be ignored.  Suppose that a molecule
diffuses in the fluid and is absorbed whenever the molecule touches the
surface of any of the traps or the wall of the beaker.  Our basic goal is to
compute the probability that the molecule is absorbed by the traps or by the
beaker wall.  In the latter case, the diffusing molecule can be viewed as
escaping from the beaker.  This type of system arises naturally in the
absorption of photons in a multiple scattering medium, where extensive
literature has discussed absorption and escape (see, e.g.,
\cite{BNHW87,CPW90,CMZ97,W98,W00,book:light,cloak}).

\begin{figure}[ht]
\begin{center}
  \label{beaker}
\includegraphics[width=0.275\textwidth]{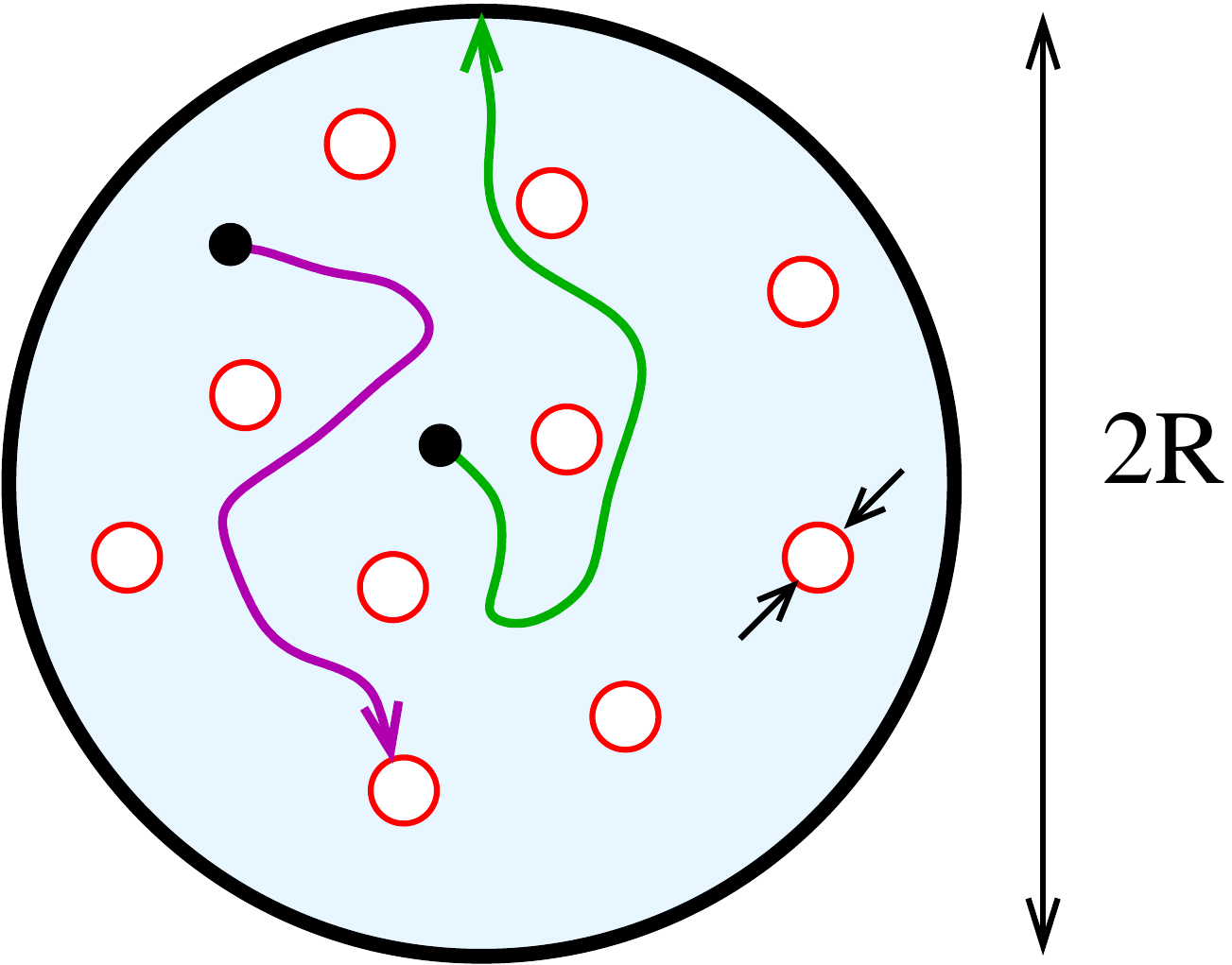} 
\caption{Sketch of a fluid-filled spherical beaker of radius $R$ that
  contains reactive traps of radius $a$.  Shown are two diffusing molecules,
  one that escapes the beaker (green) and one that is absorbed by one of the
  traps (magenta).}
\end{center}
\end{figure}

For simplicity, we assume that the traps are identical spherical particles
whose radii $a>0$ are small compared to the beaker radius.  To simplify
matters, we view the traps as immobile. This is a reasonable assumption if
the traps are much larger than the diffusing molecules, and the treatment of
the general case is essentially identical as we will discuss.  We assume that
the number of traps is sufficiently large that fluctuations in their local
density can be ignored.  In this limit, it is generally possible to replace
the discrete traps by an effective average trapping medium.  The dynamics of
the molecules can then be described by a reaction-diffusion equation, for
which explicit solutions can be obtained by standard methods.  The converse
situation where the beaker contains a few traps or the trapping efficiency of
each trap is distributed is challenging, because the trapping probability
depends in detail on the trap positions, and the effective-medium approach no
longer applies.

Eventually all the molecules are absorbed by either the traps or by the
beaker wall.  What is the fraction $E$ that reach the beaker wall?  What is
the fraction $T$ that get absorbed by the traps? Our goal is to calculate
these escape and trapping probabilities, $E$ and $T$ respectively, for simple
geometries.  We will first study the case of a spherical beaker of radius $R$
in three dimensions and then generalize to arbitrary spatial dimensions $d$
and to beakers of other simple shapes, such as a parallelepiped and a
cylinder.

We begin, in Sec.~\ref{formal} by writing the dynamical equation and governs
the escape and trapping probabilities and develop an equivalent
time-independent description for these probabilities.  Although the molecule
density evolves with time, it is possible to recast the infinite-time escape
and trapping probabilities as a simpler time-independent problem.  In
Sec.~\ref{single}, we use this perspective to solve the idealized situation
of a single trap of radius $a$ at the center of a beaker of radius $R$.  We
also extend this solution to a non-concentric geometry in two dimensions.

In Sec.~\ref{rra}, we treat our primary example of the escape of a diffusing
molecule from a spherical beaker that contains $N\gg 1$ traps.  In principle,
we have to solve the diffusion equation in the available volume---the region
exterior to the traps and interior to the beaker---and then compute the time
integrated fluxes to the traps and to the beaker walls.  A full solution to
this formidable problem is not feasible, and instead we apply the powerful
reaction-rate approach (RRA)~\cite{smol,chandra,R01,book,ovchin,B93,gleb}.
Here we replace the discrete traps by an spatially uniform effective trapping
medium, leading to a problem that is readily soluble by standard methods.

Generically, one might anticipate that the exit probability should depend on
two dimensionless parameters: $N$ and $a/R$.  We shall show, however, that
the exit probability $E(N,a/R)$ depends only on a {\em single} parameter,
which is the product of $N$ and $a/R$:
\begin{equation}
\label{lambda} 
\lambda=N\,\frac{a}{R}\,.
\end{equation}
If one knew in advance that the behavior is determined by a single parameter,
one might anticipate that it might be $N(a/R)^3$, the ratio of the trap
volume to the system volume, or perhaps $N(a/R)^2$, the ratio of the total
surface of the traps to the total exit area. Thus the dependence on a single
parameter, specifically on $\lambda=Na/R$, could be puzzling at first sight;
we shall show, however, that this dependence follows naturally from the
structure of the RRA.

In Sec.~\ref{der}, we obtain the explicit solution for the exit probability
$E(\lambda)$ for a spherical beaker in three dimensions by the RRA.  The exit
probability is given by the following compact formula
\begin{equation}
\label{Exit:neat}
E(\lambda) = 3\,\frac{\coth(\sqrt{3\lambda})}{\sqrt{3\lambda}}-\frac{1}{\lambda}
\end{equation}
The form of the exit probability is sensitive to the shape of the beaker, but
the limiting behaviors
\begin{align}
\label{asymp}
\begin{split}
T&=1-E\sim \lambda \qquad \text{as\ } \lambda\to 0, \\
E&\sim \lambda^{-1/2} \qquad\qquad \text{as\ } \lambda\to \infty,
\end{split}
\end{align}
are universal, they generally apply for non-pathological beaker shapes; only
the amplitudes of these scaling laws are shape dependent.  It bears
emphasizing that the volume fraction occupied by the traps $\varphi=N(a/R)^3$
is generally small in realistic systems.  However, since
$\lambda=N^{2/3}\,\varphi^{1/3}$, the parameter $\lambda$ does not have to be
small, even for $\varphi\ll 1$.  Thus a dilute system of traps can be
strongly absorbing.  A similar situation arises in the case of small
absorbing receptors on the surface of an otherwise reflecting
sphere~\cite{BP77}.  This system can have an absorption coefficient that is
nearly the same as that of a perfectly absorbing sphere for a small receptor
density.

\section{Time-Independent Formalism}
\label{formal}

We seek the exit probability $E({\bf r}_0)$ for a molecule to be absorbed at
the surface of the beaker, given that the molecule starts at ${\bf r}_0$.
There is also the complementary trapping probability $T({\bf r}_0)$ for a
molecule to be absorbed at the surface of one of the traps.  To determine the
exit probability, we should solve the diffusion equation
\begin{equation}
\label{dif} 
\frac{\partial P}{\partial t}=D\nabla^2 P
\end{equation}
for the molecular density $\rho({\bf r},t)$, with the initial condition
$\rho({\bf r},t\!=\!0)=\delta({\bf r}-{\bf r}_0)$, and with absorbing
boundary conditions $\rho=0$ on the surfaces of all traps and on the beaker
wall.  The exit probability $E({\bf r}_0)$ equals the local flux
$-D\nabla \rho$, integrated over the beaker surface and over all time.  If we
assume that the molecules are uniformly distributed throughout the beaker,
the exit probability, averaged over the initial molecular positions, is
\begin{equation}
\label{T-def} 
E=\frac{\int d{\bf r}_0\,E({\bf r}_0)}{\int d{\bf r}_0}\,,
\end{equation}
where the integrals extend over the allowed spatial region of the molecules.

Because the time does not appear in $E({\bf r}_0)$, the problem can be recast
into a time-independent form by integrating Eq.~\eqref{dif} over all time.
By this approach, one finds that $E({\bf r}_0)$ satisfies the backward
Kolmogorov equation~\cite{R01,book}
\begin{equation}
\label{Lap}
\nabla_0^2 E=0,
\end{equation}
where the subscript $0$ indicates the the derivatives in the Laplacian refer
to the initial coordinates of the molecule.  This backward equation should be
solved subject to the boundary conditions
\begin{equation*}
E|_{\rm traps}=0, \qquad\qquad E|_{\rm wall}=1\,.
\end{equation*}
The first condition merely states that a molecule starting on a trap surface
cannot reach the beaker wall, while the second states a molecule starting on
the beaker wall necessarily exits.  The exit probability thus satisfies the
Laplace equation for the electrostatic potential in the same geometry, with
the traps and the beaker wall substituted by conductors.  We may now utilize
well-known results for the corresponding electrostatic system \cite{S,LL} to determine
the exit probability.

\section{Single Trap}
\label{single}

We first determine the exit probability $E$ for a single spherical trap of
radius $a$ centered inside a spherical beaker of radius $R$.  For ease of
notation, we define $E_-(r)$ be the probability that a molecule that starts
at distance $r$ from the origin will be absorbed by the inner sphere (the
trap); thus $E_-(r)$ corresponds to the trapping probability defined above.
Similarly, let $E_+(r)$ be the probability to first reach the outer
sphere---the escape probability defined previously.  As discussed above,
$E_+$ and $E_-$ satisfy the Laplace equation
\begin{equation}
\label{Laplace}
\nabla^2 E_\pm(r)=0\,,
\end{equation}
subject to the boundary conditions
\begin{align}
\label{BC}
&E_+(a)=0, \qquad E_+(R)=1,\nonumber\\
&E_-(a)=1, \qquad E_-(R)=0.\nonumber
\end{align}
By spherical symmetry, we only need to solve the radial part of the Laplace
equation \eqref{Laplace} subject to these boundary conditions, from which the
exit probability is~\cite{R01}:
\begin{equation}
\label{Er_single}
E_+(r)=
\begin{cases}
{\displaystyle \frac{1-(a/r)^{d-2}}{1-\lambda^{d-2}}}&\quad d\ne 2,\\ \\
{\displaystyle \frac{\ln (r/a)}{\ln(1/\lambda)}}& \quad d=2,
\end{cases}
\end{equation}
with $\lambda=a/R$.  Since there is only a single ``trap'', the quantity
$\lambda$ defined here is equivalent to that given in Eq.~\eqref{lambda}.
The trapping probability is complementary quantity: $E_-(r)=1-E_+(r)$.

To compute to exit probability averaged over all molecules, we assume that
they are uniformly distributed in the annulus $a<r<R$, so that the density in
the range $(r, r+dr)$ is proportional to $r^{d-1} dr$.  The total exit
probability is then
\begin{equation}
\label{E_single}
E=\int_{a}^{R}dr\,r^{d-1}\,E_+(r)\Bigg/ \left(\int_{a}^{R}dr\,r^{d-1}\right)\,.
\end{equation}
Using \eqref{Er_single}--\eqref{E_single} gives, in three dimensions,
\begin{subequations}
\begin{equation}
\label{E_3d} 
E = \frac{1+\lambda/2}{1+\lambda+\lambda^2}\,, \qquad \lambda=\frac{a}{R}~.
\end{equation}
Notice that the fraction of molecules that exit from the beaker decreases
from 1 to 1/2 as $\lambda$ increases from 0 to 1.  When $\lambda\to 0$, the
trap at the origin has negligible size so that nearly all molecules will
escape the beaker.  Conversely, as $\lambda\to 1$, the physical region
becomes an infinitesimally thin annulus, which can be approximated as a slab.
In this case, the molecule is equally likely to escape via either boundary.
In two dimensions, the solution to the Laplace equation \eqref{Laplace} is
again straightforward to obtain and the corresponding result for the exit
probability is
\begin{equation}
\label{E_2d} 
E = \frac{1}{1-\lambda^2} - \frac{1}{2\ln(1/\lambda)}~.
\end{equation}
\end{subequations}
This expression has the same limiting behaviors as in the case of $d\ne 2$.

\begin{figure}
\centering
\includegraphics[width=0.4\textwidth]{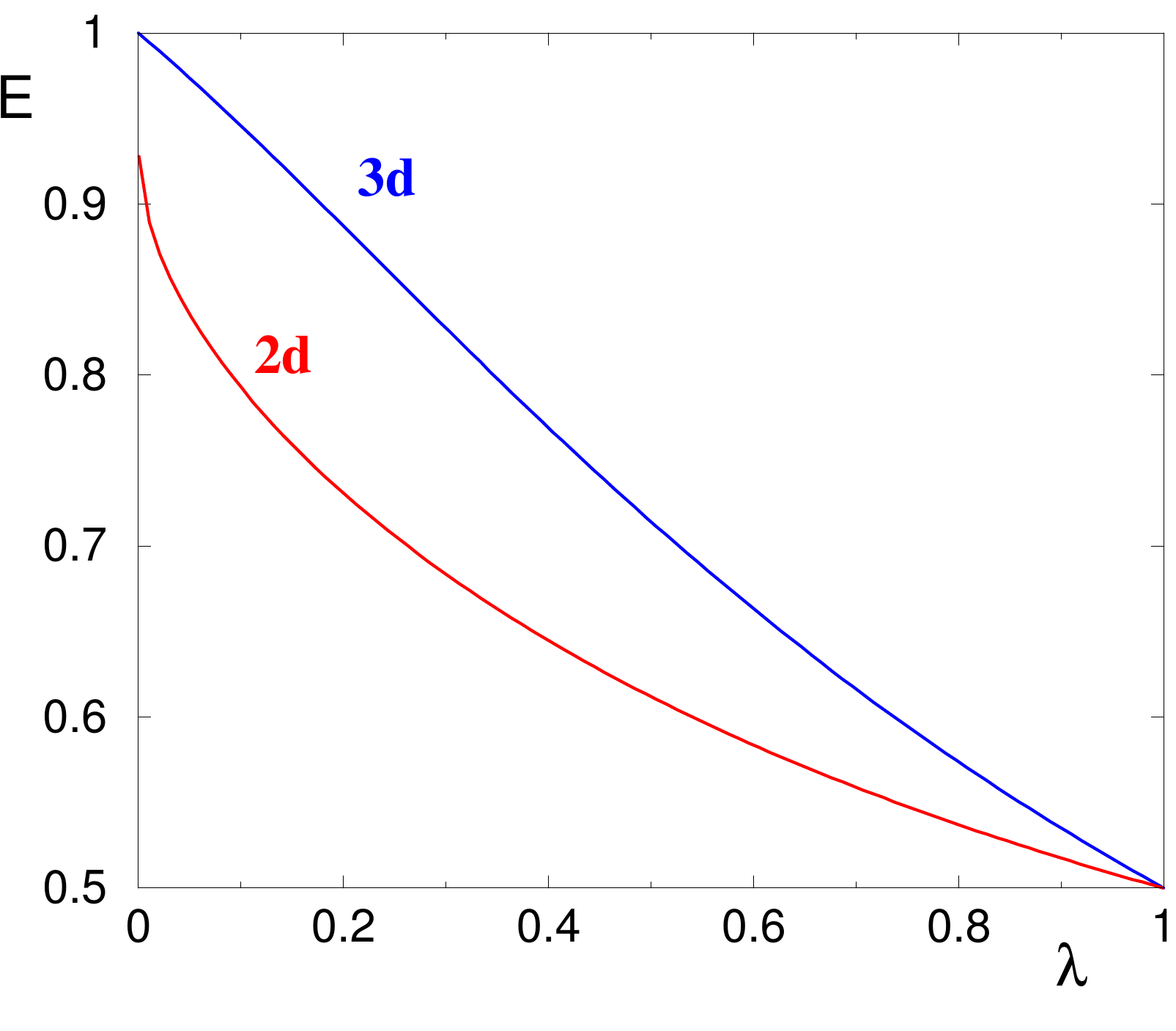}
\caption{The exit probability $E(\lambda)$, with $\lambda=R_-/R_+$ in three
  and two dimensions (Eqs.~\eqref{E_3d} and \eqref{E_2d} for a single trap
inside a beaker when the trap and beaker are concentric spheres of radii
$R_-$ and $R_+$. }
\label{Fig:EE}
\end{figure}

We can also extend the above  approach to a trap that is located
non-symmetrically within a spherical beaker.  In this case, the Laplace
equation $\nabla^2 E({\bf r})=0$ does not have a compact closed-form solution
in three dimensions~\cite{S}.  In two dimensions, however, the corresponding
solution for $E(x,y)$ is known \cite{LL}.  Suppose that the beaker is
centered at the origin while the trap is centered at $(c,0)$ (with $c<R - a$
to ensure that the trap is entirely inside the beaker).  In the corresponding
electrostatic problem, the potential between the conductors is identical to
the potential induced by two equal and opposite point charges that are
located at $(c+d_1,0)$ and $(d_2,0)$, with
\begin{equation*}
d_1(d_2-c) = a^2, \quad d_2(c-d_1) = R^2\,.
\end{equation*}
We now let
\begin{equation*}
r_1 = \sqrt{(x-c-d_1)^2+y^2}\,, \quad r_2 = \sqrt{(x-d_1)^2+y^2}
\end{equation*}
be the distances between each charge and the diffusing molecule, with the
molecule initially at $(x,y)\in \mathcal{R}$, with
\begin{equation*}
\label{region}
\mathcal{R}=\{(x,y)| x^2+y^2<R^2, ~~(x-c)^2+y^2>a^2\} 
\end{equation*}
the region between the conductors.  The probability that a molecule that
starts at $(x,y)$ is absorbed by the beaker wall is
\begin{equation}
\label{E_shift}
E(x,y)={\ln\left(\frac{r_2}{r_1}\frac{d_1}{a}\right)}\Big/{\ln\left(\frac{d_1d_2}{aR}\right)}~.
\end{equation}
The exit probability is the spatial average of $E(x,y)$ over the region
$\mathcal{R}$:
\begin{equation}
E = \frac{1}{\pi(R^2 - a^2)}\int\!\!\!\int_{\mathcal{R}} dx dy\, E(x,y)\,.
\end{equation}
This double integral does not seem to be expressible through known functions,
however.  Thus for the simplest setting of a single circular trap inside a
circular beaker, the exit probability has a closed form only when these two
objects are concentric.

\section{Reaction-Rate Approach} 
\label{rra}

The Smoluchowski reaction-rate approach (RRA) is an effective-medium
approximation that replaces the complicated influences of each individual
trap by their average influence (see, e.g.,
\cite{R01,book,smol,chandra,ovchin,B93,gleb}).  In this approach, the
diffusion equation \eqref{dif} is replaced by the reaction-diffusion equation
\begin{equation}
\label{dr} 
\frac{\partial \rho}{\partial t}=D\nabla^2 \rho - K\rho\,,
\end{equation}
where $K$ is the reaction rate that accounts for the average absorption rate
of diffusing molecules by the traps.  For identical spherical traps, the
reaction rate is~\cite{R01,book,smol,chandra,ovchin,B93,gleb})
\begin{subequations}
\begin{equation}
\label{K} 
K=4\pi Da\,n\,,
\end{equation}
where $n=N/V$ (with $N$ the number of traps and $V$ the beaker volume) is the
trap density.  This expression for $K$ is asymptotically exact when the
volume fraction of the traps is small, so that their interactions can be
neglected.  The RRA can be generalized in a straightforward way to cases
where the traps are mobile and where the size of the molecules cannot be
ignored.

It is easiest to compute the reaction rate via the electrostatic analogy
(apparently first noticed in the classical paper~\cite{BP77};
see~\cite{B93,book} for reviews).  This approach gives
\begin{equation}
\label{Kc} 
K=4\pi D n C\,,
\end{equation}
\end{subequations}
where $C$ is the electrical capacitance of the trap (assuming that it is a
perfect conductor of the same shape).  As examples, a sphere of radius $a$
has capacitance $C=a$, so that \eqref{Kc} reduces to \eqref{K}; a disk of
radius $a$ has capacitance $2a/\pi$, leading to $K=8 Dan$. The electrical
capacitance of ellipsoids is also known.  For instance, for oblate spheroids
with semi-axes $a=a\leq b$ the reaction rate is
\begin{equation*}
K=4\pi Dn\,\frac{\sqrt{a^2-b^2}}{\cos^{-1}(b/a)}\,.
\end{equation*}
In the following we treat spherical traps and use \eqref{K}; arbitrary
geometries can be obtained by straightforward generalization.

Within the RRA, we need to solve the reaction-diffusion equation with an
absorbing boundary condition at the beaker wall
\begin{equation}
\label{out} 
\rho\big|_{\rm wall}=0\,,
\end{equation}
for a spatially uniform initial condition $\rho(\mathbf{r},t=0)=1$.  Since
the molecular flux to the wall is $-D\nabla \rho$, the total number of
molecules that exit the beaker is
\begin{equation*}
-\int_0^\infty dt\int dS\,{\bf n}\cdot D\nabla \rho\,,
\end{equation*}
where the second integral is over the beaker wall and ${\bf n}$ is the unit
normal to the wall.  Dividing by the total initial number of molecules $V$
gives the exit probability
\begin{equation}
\label{exit-gen} 
E=\frac{1}{V}\int_0^\infty dt\int dS\,{\bf n}\cdot D\nabla \rho\,.
\end{equation}

\section{Trapping and Escape Probabilities}
\label{der}
We now compute the escape and trapping probabilities for: (i) a spherical
beaker in three directions, (ii) a spherical beaker in general dimensions,
and finally (iii) other simple beaker shapes in three dimensions.
Throughout, we consider the traps to be identical small spheres; however, our
results can be extended to traps of other shapes if the electrostatic
capacitance of this shape is known.

\subsection{Three dimensions}

For spherical traps, the reaction rate is given by~\eqref{K}.  When the
beaker is a ball of radius $R$, the surface integral in (\ref{exit-gen}) is
readily performed and the exit probability is
\begin{equation}
\label{E-ball}
E=\frac{3D}{R}\int_0^\infty dt
\left[-\frac{\partial \rho}{\partial r}\right]\Bigg|_{r=R}~.
\end{equation}
To compute this integral, we must solve the reaction-diffusion equation
\eqref{dr}.  This solution (Appendix~\ref{3d}) gives the concentration $\rho$
as the Fourier series
\begin{equation}
\label{rho3}
\rho(r,t)=
\frac{2}{r}\sum_{n=1}^\infty \frac{(-1)^{n+1}}{\pi n}\,
e^{-(n^2\pi^2+3\lambda)t}\,\,\sin(n \pi r)\,.
\end{equation}
Substituting this expression in \eqref{E-ball} and performing the integral,
the exit probability is
\begin{equation}
\label{Exit}
E=\sum_{n=1}^\infty\frac{6}{n^2\pi^2+3\lambda}
=3\,\frac{\coth(\sqrt{3\lambda})}{\sqrt{3\lambda}}-\frac{1}{\lambda}~.
\end{equation}
The second equality has been obtained by standard residue calculus
methods. 

\begin{figure}
\centering
\includegraphics[width=0.4\textwidth]{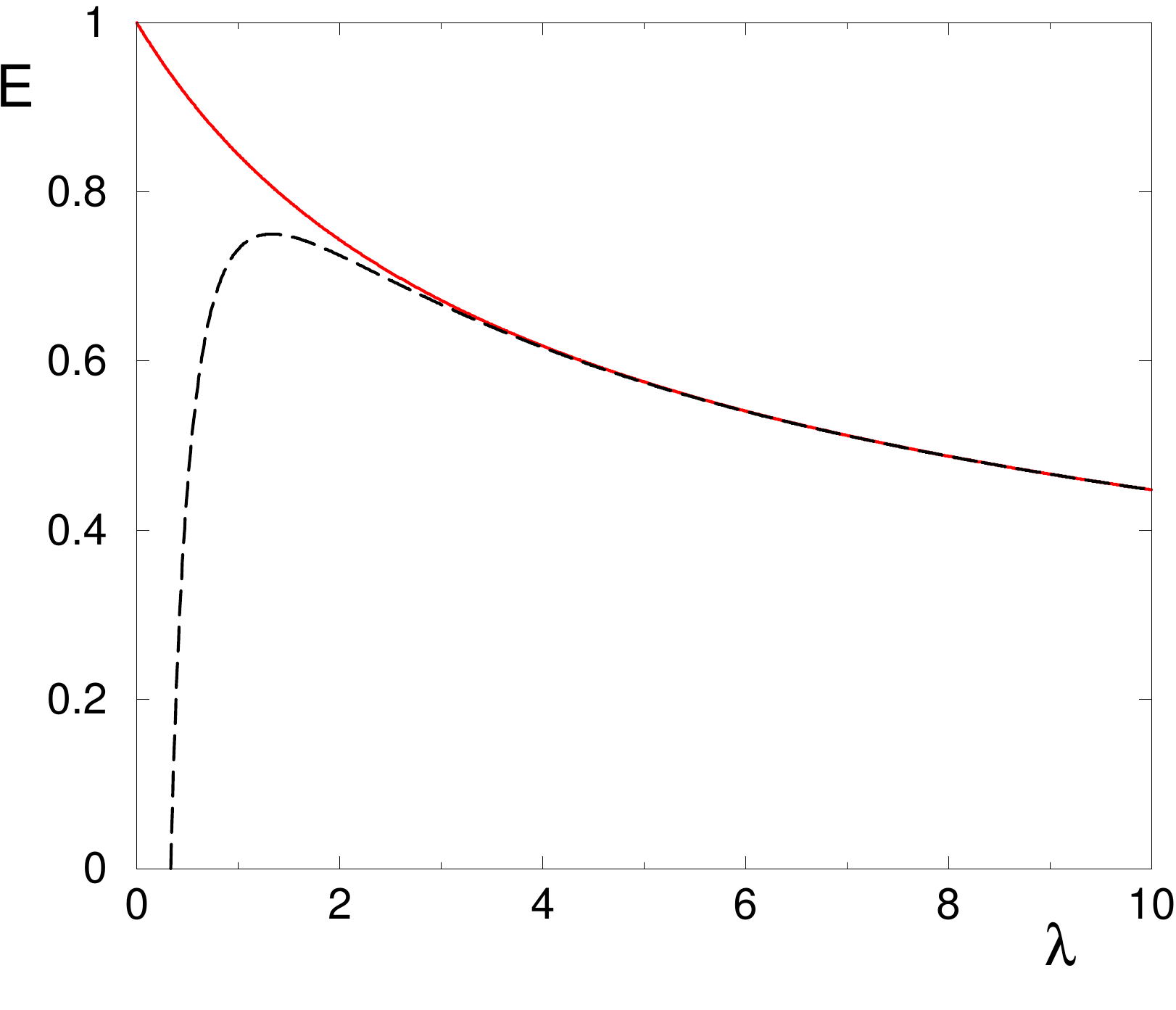}
\caption{The exit probability $E(\lambda)$, with $\lambda$ given by
  \eqref{lambda} when the number of traps is large and the beaker is a ball
  of radius $R$. The solid curve is the exact result Eq.~\eqref{Exit}, while
  the dashed curve is the approximation, Eq.~\eqref{Exit-large}. }
\label{Fig:Exit_EA}
\end{figure}

The limiting behaviors of Eq.~\eqref{Exit} are instructive.  As $\lambda\to
0$, we obtain
\begin{subequations}
\begin{align}
\label{Exit-large} 
E& =1-\frac{1}{5}\,\lambda+\frac{2}{35}\,\lambda^2-\frac{3}{175}\,\lambda^3
+\frac{2}{385}\,\lambda^4-\ldots
\end{align}
while in the opposite $\lambda\to \infty$ limit
\begin{align}
\label{Exit-large} 
E\simeq \sqrt{\frac{3}{\lambda}}-\frac{1}{\lambda}~.
\end{align}
\end{subequations}
The latter approximation is almost indistinguishable from the exact solution
\eqref{Exit} for $\lambda\agt 3$ (see Fig.~\ref{Fig:Exit_EA}), because the
next correction term in \eqref{Exit-large} is of the order of
$e^{-\sqrt{12\lambda}}$.  As we shall see, these two limiting behaviors of
$T\sim\lambda$ for $\lambda\to 0$ and $E\sim \lambda^{-1/2}$ for
$\lambda\to\infty$ arise quite generally.

\subsection{General dimensions}

For general spatial dimensions, we must solve the $d$-dimensional
reaction-diffusion \eqref{dr}.  Using spherical symmetry, we assume that the
initial condition is a normalized spherical shell of probability at $r=r_0$:
\begin{equation*}
\rho(r,t\!=\!0)= \frac{\delta(r-r_0)}{\Omega_d r_0^{d-1}}~,
\end{equation*}
where $\Omega_d$ is the surface area of a $d$-dimensional unit sphere.  We
now Laplace transform \eqref{dr} and re-express the radius in dimensionless
units $x=r\sqrt{s/D}$, after which the reaction-diffusion equation \eqref{dr}
becomes
\begin{equation}
\label{bessel}
\tilde\rho\,'' + \frac{d-1}{x}\,\tilde\rho\,' - \Big(1+\frac{\lambda}{s}\Big)\tilde\rho = -\frac{s^{(d-2)/2}}{D^{d/2}}
\frac{\delta(x-x_0)}{\Omega_d x_0^{d-1}}~.
\end{equation}
Here the tilde denotes the Laplace transform, and the prime denotes
differentiation with respect to $x$.  This equation should be solved subject
to the boundary condition $\rho(R,t)=0$.

From the solution to Eq.~\eqref{bessel} (see Appendix~\ref{gend}), the
probability $E(r_0)$ that a diffusing molecule that starts at $r_0$
ultimately escapes to the beaker wall is given by
\begin{equation}
\label{Egend}
E(r_0)=\left(\frac{r_0}{R}\right)^\nu \frac{I_\nu(\mu r_0)}{I_\nu(\mu R)}~.
\end{equation}
Here $I_\nu$ and $K_\nu$ are the modified Bessel functions of order $\nu$,
with $\nu=(2\!-\!d)/2$, and $\mu=\sqrt{1\!+\!\lambda/s}$. 

A more meaningful measure of the trapping efficiency of the system is the
escape probability when the fluid initially contains a uniform density of
diffusing molecules.  Here we need to integrate \eqref{Egend} over all radii
and also modify \eqref{Egend} for the initial condition of a uniform density
in the entire beaker, rather than a spherically symmetric density shell at
radius $r_0$.  For the latter, we merely need to multiply the result of
integrating \eqref{Egend} over all radii by the factor ${d}/{R}$ to account
for the difference is the normalizations between a uniform density within a
sphere and on a spherical surface.

For the spatial integration, we make use of the identity
\begin{equation*}
\frac{d}{dz}\big[z^\nu I_\nu(z)\big] = z^\nu I_{\nu-1}(z)\,,
\end{equation*}
as well as $I_\nu=I_{-\nu}$ to write
\begin{align}
\label{Eint-gend}
E&= \frac{d}{R}\int_0^R\! r^{d-1}\, E(r)\, dr \nonumber \\
&= \frac{d}{R} \int_0^R \! r^{d-1+\nu} \,\,\frac{I_{\nu}(\mu r)}{I_{\nu}(\mu
  R)}\,dr\nonumber \\
&= \frac{d}{R} \int_0^R\! r^{d/2} \,\,\frac{I_{(d-2)/2}(\mu r)}{I_{(d-2)/2}(\mu
  R)}\,dr\nonumber \\
&=\frac{d}{\mu R}\,\, \frac{I_{d/2}(\mu R)}{I_{(d-2)/2}(\mu R)}~.
\end{align}
In three dimensions, the above expression reduces to that given in
\eqref{Exit}.  From \eqref{Eint-gend},  the exit
probability has the same asymptotic behaviors \eqref{asymp} in all dimensions $d>2$.

\section{Discussion}
\label{sec:Discussion}

We studied the trapping characteristics of diffusing molecules that are
absorbed whenever they either reach the boundary of a finite fluid-filled
beaker or touch traps within the fluid.  The solution for a single trap is
simple for the example of a spherical trap at the center of a spherical
beaker.  It is impractical, however, to generalize this analytical approach
for more than a single trap because it involves solving the diffusion
equation subject to absorbing boundary conditions on the surface of all traps
and the surface of the beaker.  However, the problem greatly simplifies when
there are many traps.  In this limit, we can replace the traps by an
effective and uniformly absorbing medium---the reaction rate approximation
(RRA).  We showed how to compute the fraction of molecules that escape the
domain for the situation where the total number of traps is large.
Remarkably, this escape probability depends only on a single parameter
$\lambda$, the product of the total number of traps and the ratio of the
characteristic size of the trap to the characteristic size of the domain.
Our general results is that the exit probability from the beaker, $E$, exhibits 
universal asymptotic behaviors \eqref{asymp} in all dimensions greater than two.

While our model describes the absorption of diffusing molecules in a beaker
that contains a reactive fluid, there are other other potential scenarios.
For example, the domain could be a cell and the diffusing molecules may exit
the cell membrane, or may be absorbed by traps within the cell and thereby
perform some biologically useful function.  Here, exit may be treated as a
loss and one would like to estimate the magnitude of the loss.  In realistic
systems, the parameter $\lambda$ is usually large, so our results can be
interpreted as the generic claim that this loss, viz., the fraction of
molecules exiting through the cell membrane, universally scales as
$A/\sqrt{\lambda}$ when $\lambda\gg 1$.  Only the amplitude $A$ depends on
the shape of the cell and on the shape of the traps.

Within the RRA, we determined the exit and trapping probabilities without
needing to solve the underlying reaction-diffusion equation.  In addition to
simplifying the calculations, the RRA provides additional insights that would
be very difficult to obtain by direct means.  For example, for a cylindrical
beaker of radius $R$ and height $H$, the exit probability could, in
principle, depend on three dimensionless parameters: $N$, $a/R$, and $R/H$.
The RRA predicts that for very tall cylinders ($R\ll H$), the exit
probability again actually depends on the single parameter $\lambda=Na/H$.
It is worth noting that within the RRA formalism it also is straightforward
to account for diffusing traps as well as a non-zero radii for both the traps
and the molecules.  All that is needed is to amend the parameter $\lambda$
from $Na/R$ to
\begin{equation}
\label{lambda-last} 
\lambda=N\,\frac{a_m+a_t}{R}\,\frac{D_m+D_t}{D_m}\,,
\end{equation}
where the subscripts $m$ and $t$ refer to the molecules and the traps,
respectively.

\bigskip We thank Sergei Rudchenko, Nagendra K. Panduranga and Kirill Korolev
for helpful discussions.  This work was partly supported by the National
Science Foundation under grant No.~DMR-1608211.

\appendix

\section{Concentration in Three Dimensions}
\label{3d}

To solve the reaction-diffusion equation \eqref{dr} in three dimensions, we
use a simplification that transforms the radial Laplacian that operates on
the quantity $u(r)/r$ to a one-dimensional Laplacian:
\begin{equation}
\label{3-1d}
\frac{1}{r^2}\,\frac{\partial}{\partial r}
\left(r^2\,\frac{\partial}{\partial r}\right)\frac{u}{r}
= \frac{1}{r}\, \frac{\partial^2 u}{\partial r^2}~.
\end{equation}
This identity suggests that we work with $u=\rho\,r$, after which the
reaction-diffusion equation, with $K$ given by (\ref{K}), becomes
\begin{equation}
\label{dru} 
\frac{\partial u}{\partial t}=D\left(\frac{\partial^2}{\partial r^2} 
- \frac{3Na}{R^3}\right)u\,.
\end{equation}
It is useful to now introduce the dimensionless variables
\begin{equation}
\label{dimless} 
\overline{r}=\frac{r}{R}\,,\qquad
\overline{t}=\frac{Dt}{R^2}~,
\end{equation}
and then drop the overbars to simplify notation in the following.
Equation~\eqref{dru} becomes
\begin{equation}
\label{main} 
\frac{\partial u}{\partial t}=\frac{\partial^2 u}{\partial r^2} 
- 3\lambda u, \qquad \lambda=N\,\frac{a}{R}~.
\end{equation}
We want to solve \eqref{main} subject to the condition of a uniform initial
density
\begin{subequations}
\label{3C}
\begin{equation}
\label{inu} 
u|_{t=0}=r\,,
\end{equation}
an absorbing boundary condition at the beaker wall
\begin{equation}
\label{outu} 
u|_{r=1}=0\,,
\end{equation}
and the condition
\begin{equation}
\label{zerou} 
u|_{r=0}=0\,
\end{equation}
\end{subequations}
that ensures that $\rho=u/r$ is well-defined in the origin.  The exit
probability \eqref{E-ball} now becomes
\begin{equation}
\label{exit} 
E=3\int_0^\infty dt\left[-\frac{\partial u}{\partial r}\right]\Bigg|_{r=1}
\end{equation}

We use Fourier analysis (see, e.g., \cite{DMK}) to solve Eq.~\eqref{main}
subject to \eqref{3C}.  Here it is useful to extend the domain of $u(r,t)$
from $[0,1]$ to $[-1,1]$.  For convenience we also extend $u$ to be odd
(i.e., $u(-r)=-u(r)$), so that the boundary condition (\ref{zerou})
manifestly holds.  Then the Fourier expansion of $u(r,t)$ contains only sine
terms:
\begin{equation}
\label{Fourier} 
u(r,t)=\sum_{n=1}^\infty F_n(t)\,\sin(n\pi r)\,.
\end{equation}
From the initial condition (\ref{inu}) we obtain
\begin{equation*}
F_n(0)=2\int_0^1 dr\,r\,\sin(n\pi r)=2\,\frac{(-1)^{n+1}}{\pi n}~.
\end{equation*}
Substituting (\ref{Fourier}) into (\ref{main}) gives the differential
equation for the amplitudes
\begin{equation*}
\frac{d F_n}{dt}=-(n^2\pi^2+3\lambda)\,F_n\,.
\end{equation*}
Solving this equation, the concentration $\rho(r,t)$ may be written as the
infinite series in Eq.~\eqref{rho3}.

\section{General Dimensions}
\label{gend}

The Laplace transform of the reaction-diffusion equation Eq.~\eqref{bessel} is
\begin{equation}
\label{Abessel}
\tilde\rho'' + \frac{d-1}{x}\,\tilde\rho' - 
\Big(1+\frac{\lambda}{s}\Big)\tilde\rho = -\frac{s^{(d-2)/2}}{D^{d/2}}
\frac{\delta(x-x_0)}{\Omega_d x_0^{d-1}}~,
\end{equation}
where $x=r\sqrt{s/D}$ is the scaled coordinate, and $x$ is in the range
$(0,X)$, with $X=R\sqrt{s/D}$.  This equation should be solved subject to the
absorbing boundary condition at $x=X$.  The left-hand side of \eqref{Abessel}
is a Bessel differential equation that we solve separately for $x<x_0$ and
$x>x_0$, and then patch together these two solutions by the standard joining
conditions for Green's functions.  In each subdomain $x<x_0$ and $x>x_0$, the
elemental solutions have the form
\begin{align}
\begin{split}
\label{rho}
\rho_<&= A x^\nu I_\nu(\mu x)\,,\\
\rho_>&= B x^\nu I_\nu(\mu x)+Cx^\nu K_\nu(\mu x)\,.
\end{split}
\end{align}
Here $I_\nu$ and $K_\nu$ are the modified Bessel functions of order $\nu$,
with $\nu=(2\!-\!d)/2$, and $\mu=\sqrt{1\!+\!\lambda/s}$.  The subscripts $<$
and $>$ denote the solution in the regions $x<x_0$ and $x>x_0$.  The interior
solution does not contain the function $K_\nu$ because $K_\nu$ diverges at
the origin.  Invoking the absorbing boundary condition $\rho=0$ at $x=X$, as
well as the continuity of the Green's function at $r=r_0$, the solution
\eqref{rho} can be expressed in a form that is manifestly continuous and
vanishes at $x=X$:
\begin{equation*}
\begin{split}
\label{rho-full}
\rho_<&= A x^\nu I_\nu(\mu x)\big[I_\nu(\mu x_0)K_\nu(\mu X)-I_\nu(\mu X)K_\nu(\mu x_0)\big]\,,\\
\rho_>&= A x^\nu I_\nu(\mu x_0)\big[I_\nu(\mu x)K_\nu(\mu X)-I_\nu(\mu X)K_\nu(\mu x)\big]\,.
\end{split}
\end{equation*}
The constant $A$ is determined by the joining condition
\begin{equation*}
D(\rho_>'-\rho_<')\Big|_{x_0} = -
\frac{s^{(d-2)/2}\,\delta(x-x_0)}{D^{d/2}\,\Omega_d\,\, x_0^{d-1}}~.
\end{equation*}
For differentiating the Green's function, we use
\begin{equation*}
I_\nu'=\frac{\nu}{x}I_\nu+I_{\nu-1}\quad \text{and}\quad
K_\nu'=\frac{\nu}{x}K_\nu-K_{\nu-1}\,,
\end{equation*}
while for $\rho_>'-\rho_<'$, we  use the Wronskian relation
\begin{equation*}
x_0\big[I_{\nu-1}(x_0)K_\nu(x_0)+K_{\nu-1}(x_0)I_\nu(x_0)\big]=1.
\end{equation*}

With these identities and performing some tedious but straightforward
algebra, the amplitude of the Green's function is given by
\begin{equation*}
A= \frac{(s/x_0)^{(d-2)/2}}{D^{1+d/2}\,\Omega_d\,I_\nu(\mu X)}\,.
\end{equation*}
Finally, we compute the flux $-D\rho'$ to the outer boundary, integrate over
the surface of the sphere, and then take the $s\to 0$ limit of the Laplace
transform to obtain the probability $E(r_0)$ that a diffusing molecule that
starts at $x_0$ ultimately escapes to the beaker wall.  This is
Eq.~\eqref{Egend}.

\section{Other Geometries} 
\label{ap}

Here we solve (\ref{dr}) and compute the exit probability for two additional
cases.

\subsection{Parallelepiped} 

Consider the parallelepiped of size $L_1\times L_2\times L_3$.  The
reaction-diffusion equation is
\begin{equation}
\label{dr-rec} 
\frac{\partial \rho}{\partial t}=D\left(\nabla^2 \rho - 
\frac{4\pi a N}{L_1\,L_2\,L_3}\,\rho\right)~.
\end{equation}
We seek a solution in the form of a Fourier sine series
\begin{equation}
\label{F-rec} 
\rho(r,t)=\sum_{n_1,n_2,n_3\geq 1} F_{\bf n}(t)\prod_{j=1}^3
\sin\left( n_j\,\pi\frac{x_j}{L_j}\right)~,
\end{equation}
which ensures that the absorbing boundary condition \eqref{out} is
automatically satisfied on the walls $x_j=0,L_j$.  Substituting \eqref{F-rec}
into \eqref{dr-rec} and solving for $F_{\bf n}(t)$ gives
\begin{equation*}
F_{\bf n}(t)=F_{\bf n}(0)\,\exp[-\pi^2\Lambda({\bf n}) Dt]\,,
\end{equation*}
where we use the shorthand notation
\begin{equation*}
\Lambda({\bf n})=\left(\frac{n_1}{L_1}\right)^2
+\left(\frac{n_2}{L_2}\right)^2
+\left(\frac{n_3}{L_3}\right)^2
+\frac{4}{\pi}\,\frac{a N}{L_1\,L_2\,L_3}\,.
\end{equation*}
For the spatially uniform initial condition we find that $F_{\bf n}(0)=0$ if
at least one index $n_j$ is even; when all $n_j$'s are odd, we have
\begin{equation*}
F_{\bf n}(0)=\frac{8^2}{\pi^3}\,\frac{1}{n_1\,n_2\,n_3}\,.
\end{equation*}
After some straightforward calculations, the exit probability is
\begin{equation*}
E=\frac{8^3}{\pi^6}\sum_{\bf n}
\left[\frac{1}{L_1^2\,n_2^2\,n_3^2}
+\frac{1}{L_2^2\,n_3^2\,n_1^2}
+\frac{1}{L_3^2\,n_1^2\,n_2^2}\right]\frac{1}{\Lambda({\bf n})}~.
\end{equation*}
Here and below the $\sum_{\bf n}$ runs over $n_1,n_2,n_3$ which are all
positive and odd.

Let us consider two examples in more detail.  For the cube with
$L_1=L_2=L_3=L$, we obtain
\begin{equation}
\label{exit_cube}
E=\frac{8^3}{\pi^6}\sum_{\bf n}
\frac{1}{n_1^2\,n_2^2\,n_3^2}\,\,\frac{{\bf n}^2}{{\bf n}^2+\lambda} \,,
\end{equation}
where we use the shorthand ${\bf n}^2=n_1^2+n_2^2+n_3^2$ with
$\lambda=4aN/\pi L$~\cite{note}.
Using \eqref{exit_cube} and the identity~\cite{knuth}
\begin{equation}
\label{Euler}
\sum_{n~\text{odd}} \frac{1}{n^2} = \frac{\pi^2}{8}\,,
\end{equation}
we determine the trapping probability 
\begin{equation}
\label{trapping_cube}
T=\frac{8^3}{\pi^6}\sum_{\bf n}
\frac{1}{n_1^2\,n_2^2\,n_3^2}\,\,\frac{\lambda}{{\bf n}^2+\lambda} ~.
\end{equation}
The small $\lambda$ expansion of this trapping probability is
\begin{equation}
\label{trapping_exp}
T=A_1\lambda - A_2\lambda^2 + A_3\lambda^3-A_4\lambda^4 +\ldots\,,
\end{equation}
where
\begin{equation*}
A_p=\frac{8^3}{\pi^6}\sum_{\bf n}
\frac{1}{n_1^2\,n_2^2\,n_3^2}\,\,\frac{1}{(n_1^2+n_2^2+n_3^2)^p}~.
\end{equation*}
To determine the large-$\lambda$ behavior, we use symmetry and re-write
\eqref{exit_cube} as
\begin{equation}
\label{exit_cube_1}
E=3\,\frac{8^3}{\pi^6}\sum_{n_2,\, n_3}
\frac{1}{n_2^2\,n_3^2}\,\sum_{n_1}\frac{1}{n_1^2+n_2^2+n_3^2+\lambda} ~.
\end{equation}
Next, we replace the summation over $n_1$ by integration
\begin{equation*}
\sum_{n_1~\text{odd}}\to \frac{1}{2}\int_0^\infty dn_1 \,,
\end{equation*}
from which, we can obtain the leading asymptotic behavior in the
$\lambda\to\infty$ limit.  Computing the integral gives
\begin{equation}
\label{exit_cube_23}
E\simeq 3\,\frac{8^3}{\pi^6}\sum_{n_2,\, n_3}
\frac{1}{n_2^2\,n_3^2}\,\frac{\pi/4}{\sqrt{n_2^2+n_3^2+\lambda}} ~.
\end{equation}
In the $\lambda\to\infty$ limit, we make the replacement
$\sqrt{n_2^2+n_3^2+\lambda}\to \sqrt{\lambda}$ and use the identity
\eqref{Euler} to find
\begin{eqnarray}
\label{exit_cube_large}
E&\simeq& 3\,\frac{8^3}{\pi^6}\sum_{n_2,\, n_3}\,\,
\frac{1}{n_2^2\,n_3^2}\,\frac{\pi}{4\sqrt{\lambda}}\,, \nonumber\\
 &=&  3\,\frac{8^3}{\pi^6} \left(\frac{\pi^2}{8}\right)^2\frac{\pi}{4\sqrt{\lambda}} \nonumber\\
& = &\frac{6/\pi}{\sqrt{\lambda}} = 3\,\sqrt{\frac{L}{\pi aN}}~.
\end{eqnarray}

As another example, consider the bar with dimensions $L_1=L_2\ll L_3$.  We
perform the summation over $n_3$ by using identity \eqref{Euler}.  Then the
exit probability becomes
\begin{equation}
\label{exit_bar}
E=\frac{8^2}{\pi^4}\sum_{n_1,n_2}
\frac{1}{n_1^2\,n_2^2}\,\,\frac{n_1^2+n_2^2}{n_1^2+n_2^2+\lambda}\,, \quad
\lambda=\frac{4}{\pi}\,\frac{a N}{L_3}\,,
\end{equation}
where the summation runs over $n_1,n_2$ which are both positive and odd.

We obtain the asymptotic behaviors following similar steps as for the cube.
The small-$\lambda$ expansion of the trapping probability is given by
\eqref{trapping_exp} with
\begin{equation*}
A_p=\frac{8^2}{\pi^4}\sum_{n_1,\, n_2}
\frac{1}{n_1^2\,n_2^2}\,\,\frac{1}{(n_1^2+n_2^2)^p}~.
\end{equation*}
Using \eqref{exit_bar} and calculations similar to
\eqref{exit_cube_23}--\eqref{exit_cube_large} we obtain
\begin{equation}
\label{exit_bar_large}
E\simeq \frac{4/\pi}{\sqrt{\lambda}} = 2\,\sqrt{\frac{L_3}{\pi aN}}~.
\end{equation}

Generally for the $L_1\times L_2\times L_3$ parallelepiped, the exit
probability decays (for $N\gg L/a$) as
\begin{equation}
E\simeq \sqrt{\frac{L_1 L_2 L_3}{\pi aN}}\left[\frac{1}{L_1}+\frac{1}{L_2}+\frac{1}{L_3}\right]~.
\end{equation}
From this general result one can recover \eqref{exit_cube_large} for the cube
and \eqref{exit_bar_large} for the bar.

\subsection{Cylinder} 

If the cylinder height $H$ greatly exceeds its radius $R$, the problem
becomes two dimensional. In the dimensionless variables of \eqref{dimless},
the reaction-diffusion equation \eqref{dr} becomes
\begin{equation*}
\frac{\partial \rho}{\partial t}=\frac{\partial^2 \rho}{\partial r^2} 
+\frac{1}{r}\,\frac{\partial \rho}{\partial r} 
-\lambda\rho, \qquad \lambda=\frac{4 a N}{H}~.
\end{equation*}
The solution is the Bessel series
\begin{equation*} 
\rho(r,t)=\sum_{n=1}^\infty A_n\,J_0(\mu_n r)\,e^{-(\mu_n^2+\lambda)t}\,.
\end{equation*}
The absorbing boundary condition is satisfied when $0<\mu_1<\mu_2<\ldots$ are
consecutive zeros of the Bessel function $J_0(\mu_n)=0$.  For the uniform
initial condition, and using the orthogonality condition
$\int_0^1 dr\,r\,J_0(\mu_n r)\,J_0(\mu_m r)=0$ for $n\ne m$, the coefficients
$A_n$ are
\begin{equation*} 
A_n=\frac{\int_0^{\mu_n} dx\,x\,J_0(x)}{\int_0^{\mu_n} dx\,x\,[J_0(x)]^2}~,
\end{equation*}
from which the exit probability is
\begin{equation*} 
E=2\sum_{n=1}^\infty A_n\,\frac{\mu_n\,J_0'(\mu_n)}{\mu_n^2+\lambda}~.
\end{equation*}
Hence $E\propto\lambda^{-1}$ for large $\lambda$.  In contrast, for
three-dimensional systems, the exit probability always scales as
$\lambda^{-1/2}$ for $\lambda\gg 1$.


\begin{thebibliography}{99}

\bibitem{BNHW87} R. F. Bonner, R. Nossal, S. Havlin, and G. H. Weiss,
  J. Opt.\ Soc.\ A \textbf{4}, 423 (1987).
  
\bibitem{CPW90} W. F. Cheong, S. A. Prahl, and A. J. Welch, IEEE J. Quantum
  Electronics \textbf{26}, 2166 (1990). 

\bibitem{CMZ97} D. Contini, F. Martelli, and G. Zaccanti, Appl.\
  Opt. \textbf{36}, 4587 (1997).
  
\bibitem{W98} G. Weiss, Appl.\ Opt.\ \textbf{37}, 3558 (1998).

\bibitem{W00} G. Weiss and A. Gandjbakhche, Phys.\ Rev.\ E \textbf{61}, 6958
  (2000).

\bibitem{book:light} F. Martelli, S. Del Bianco, A. Ismaelli and G. Zaccanti,
  {\it Light Propagation through Biological Tissue and Other Diffusive Media:
    Theory, Solutions and Software} (SPIE Press, Washington, USA, 2010).
  
\bibitem{cloak} R. Schittny, A. Niemeyer, F. Mayer, A. Naber, M. Kadic, and
  M. Wegener, Laser Photonics Rev.\ {\bf 10}, 282 (2016).
 
\bibitem{smol} M.~V.~Smoluchowski, Phys.\ Z.\ {\bf 17}, 557 (1916); {\it
    ibid} {\bf 17}, 585 (1916); Z.\ Phys.\ Chem.\ {\bf 92}, 129 (1917).

\bibitem{chandra} S.~Chandrasekhar, Rev.\ Mod.\ Phys.\ {\bf 15}, 1 (1943).

\bibitem{R01} S.~Redner, {\it A Guide to First-Passage Processes} (Cambridge
  University Press, New York, 2001).

\bibitem{book} P. L. Krapivsky, S. Redner and E. Ben-Naim, {\it A Kinetic
    View of Statistical Physics} (Cambridge University Press, New York,
  2010).

\bibitem{ovchin} A.~A.~Ovchinnikov, S.~F.~Timashev, and A.~A.~Belyi, {\it
    Kinetics of Diffusion Controlled Chemical Processes} (Nova Science
  Publishers, Commack, New York, 1989).

\bibitem{B93} H. C. Berg, {\it Random Walks in Biology} (Princeton University
  Press, Princeton, NJ, 1993).

\bibitem{gleb} G.~Oshanin, M.~Moreau, and S.~Burlatsky, Adv.\ Colloid
  Interface Sci.\ {\bf 49}, 1 (1994).

\bibitem{BP77} H.~C.~Berg and E.~M.~Purcell, Biophys.\ J. {\bf 20}, 193
  (1977).

\bibitem{S} W. R. Smythe, {\it Static and Dynamic Electricity}, $2^{\rm nd}$
  ed.\ (McGraw-Hill, New York, 1950).

\bibitem{LL} L. D. Landau and E. M. Lifshitz, {\it Electrodynamics of
    Continuous Media}, 2nd ed.\ (Pergamon, New York, 1984).

\bibitem{DMK} H.~Dym and H.~P.~McKean, {\it Fourier Series and Integrals}
  (Academic Press, New York, 1972).

\bibitem{note} We define $\lambda$ as $aN$ divided by a characteristic length
  for the geometry, and therefore $\lambda$ is slightly different for the
  sphere, cube, cylinder, etc.

\bibitem{knuth} R.~L.~Graham, D.~E.~Knuth, and O.~Patashnik, {\it Concrete
    Mathematics: A Foundation for Computer Science} (Addison-Wesley, Reading
  MA, 1989).

\end{thebibliography}
\end{document}